\documentclass[twocolumn,showpacs,preprintnumbers,amsmath,amssymb]{revtex4}

\usepackage{graphicx,epsfig}
\usepackage{dcolumn}
\usepackage{bm}

\begin{document}

\preprint{}

\title{Dynamical Mechanisms in Multi-agent Systems:
Minority Games}

\author{K. Y. Michael Wong}
\author{S. W. Lim}
\author{Zhuo Gao}

\affiliation{%
Department of Physics, Hong Kong University of Science and Technology, 
Clear Water Bay, Hong Kong, China
}%

\date{\today}

\begin{abstract}
We consider a version of large population games 
whose agents compete for resources 
using strategies with adaptable preferences. 
Diversity among the agents reduces their maladpative behavior. 
We find interesting scaling relations with diversity 
for the variance of decisions.
When diversity increases, 
the scaling dynamics is modified 
by kinetic sampling and waiting mechanisms.
\end{abstract}

\maketitle


Many natural and artificial systems involve interacting agents, 
each making independent decisions to compete for limited resources,
but globally exhibit coordinated behavior through their mutual adaptation 
\cite{global,minority}.
Examples include the competition of predators in ecology, 
buyers or sellers in economic markets, 
and routers in computer networks. 
While a standard approach is to analyse the steady state behavior 
of the system described by the Nash equilibria \cite{game},
it is interesting to consider 
the dynamics of how the steady state is approached.
Dynamical studies are especially relevant 
when one considers the effects of changing enviroment,
such as those in economics or distributed control.

The recently proposed Minority Games (MG) 
are prototypes of such multi-agent systems \cite{minority}.
The dynamical nature of the adaptive processes 
is revealed when the complexity of the agents is low,
where the final states depend on the initial conditions
\cite{statmech,garrahan}.
Here, the system exhibits large fluctuations, 
which are caused by
the initially zero preference of strategies for all agents.
However, when the game is used to model economic systems,
it is not realistic to expect that all agents enter the market
with the same preference.
Besides, in games which use public information only, 
this imply that different agents 
would maintain identical preferences of strategies 
at all subsequent steps,
which is again unlikely.
Furthermore, when the game is used to model distributed control 
in multi-agent systems,
identical preferences of strategies of the agents lead to 
{\it maladaptative} behavior, 
which refers to the bursts of the population's decisions 
due to their premature rush to certain state,
compromising the system efficiency \cite{savit,johnson}.
There were attempts of improvement
by introducing thermal noise \cite{thermal},
biased starts \cite{coolen,bias}, bias strategies \cite{yip}, 
and random initial conditions \cite{garrahan}.
However, no systematic studies have been made.

In this Letter, we consider the effects of introducing randomness 
in the initial preferences of strategies among the agents, 
focusing on the regime of low complexity, 
where analyses assuming vanishing step sizes 
are not applicable \cite{statmech,coolen}. 
Concretely, we consider a population of $N$ agents 
competing in an environment of limited resources,
$N$ being odd \cite{minority}. 
Each agent makes a decision $+$ or $-$ at each time step,
and the minority group wins.
The decisions of each agent are prescribed by {\it strategies},
which are Boolean functions
mapping the {\it history} of the winning bits
in the most recent $m$ steps
to decisions $+$ or $-$.
Before the game starts,
each agent randomly picks $s$ strategies.
Out of her $s$ strategies,
each agent makes decisions according to the most successful one
at each step; 
the success of a strategy is measured by its {\it virtual point},
which increases (decreases) by $1$ 
if it indicates a winning (losing) decision at a time step.

In contrast to early versions of the game,
the agents may enter the game with diverse preferences of their strategies.
This is done by randomly assigning $R$ virtual points 
to the $s$ strategies of each agent before the game starts.
Hence the initial virtual point of each strategy 
obeys a multinomial distribution with mean $R/s$ and variance $R(s-1)/s^2$.
The ratio $\rho\equiv R/N$ is referred to as the {\it diversity}.
In particular, for $s=2$ and odd $R$,
no two strategies have the same virtual points throughout the game,
and the dynamics of the game is deterministic,
resulting in highly precise simulation results 
useful for refined comparison with theories. 

\begin{figure}
\centerline{\epsfig{figure=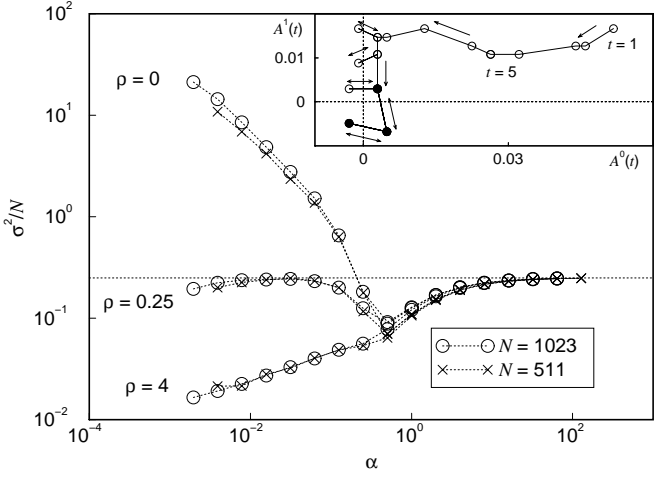,width=1\linewidth}}
\vspace{-0.4cm}
\caption{\label{va} The dependence of the variance of the population 
for decision + 
on the complexity for different diversities at $s=2$
averaged over 128 samples. 
The horizontal dotted line is the limit of random decisions.
Inset: the state motion of a sample in the phase space for $m=1$. 
Solid dots: attractor states.}
\end{figure}

To describe the macroscopic dynamics of the system,
we define the $D$-dimensional vector $A^\mu(t)$,
which is the sum of the decisions of all agents 
responding to history $\mu$ of their strategies, 
normalized by $N$, 
where $D\equiv 2^m$ is the number of histories. 
While only one of the $D$ components corresponds to the historical state 
$\mu^*(t)$ of the system, 
the augmentation to $D$ components is necessary to describe 
the attractor structure and the transient behavior of the system dynamics.
The inset of Fig. \ref{va} illustrates the convergence to the attractor 
for the visualizable case of $m=1$.
The dynamics proceeds in the direction 
which tends to reduce the magnitude of the components of $A^\mu(t)$ 
\cite{statmech}.
However, a certain amount of maladaptation always exists in the system,
so that the components of $A^\mu(t)$ overshoot,
resulting in periodic attractors with period of $2D$.
Every state $\mu$ appears as historical states 
two times in a steady-state period, 
yielding the winning bits $-$ and $+$ each exactly once.
One occurence brings $A^\mu$ from positive to negative,
and another bringing it back from negative to positive,
thus completing a cycle.
For $m=1$, 
the steady state is described by the sequence $\mu(t)=-,+,+,-$,
where both states $-$ and $+$ are followed by $-$ and $+$ once each.
For general values of $m$,
the states in an attractor 
are given by a binary de Brujin sequence of order $m+1$ \cite{deBrujin}.

As shown in Fig. \ref{va},
the variance $\sigma^2/N$ of the population for decision + 
scales as a function of the {\it complexity} $\alpha\equiv D/N$,
agreeing with previous observations \cite{savit}.
When $\alpha$ is small, games with increasing complexity 
create time series of decreasing fluctuations.
A phase transition takes place around $\alpha_c\approx 0.3$,
after which it increases gradually to the limit of random decisions, 
with $\sigma^2/N=0.25$.
When $\alpha <\alpha_c$,
we have the {\it symmetric} phase,
in which the occurences of decisions 1 and 0
responding to a given historical state $\mu$ are equal,
whereas in the {\it asymmetric} phase above $\alpha_c$,
the occurences are biased for at least some history $\mu$
\cite{symmetry}.
Figure \ref{va} also shows the data collapse for different $N$ 
for $\rho\sim 1$,
indicating that the variance is a function of $\rho$.
It is observed that the variance decreases significantly with diversity 
in the symmetric phase,
and remains unaffected in the asymmetric phase.

The dependence of the variance on the diversity is further shown
in Fig. \ref{vr} for given memory sizes $m$.
Here we focus on the physical picture of the dynamics \cite{elsewhere}.
Four regimes can be identified:

\begin{figure}
\centerline{\epsfig{figure=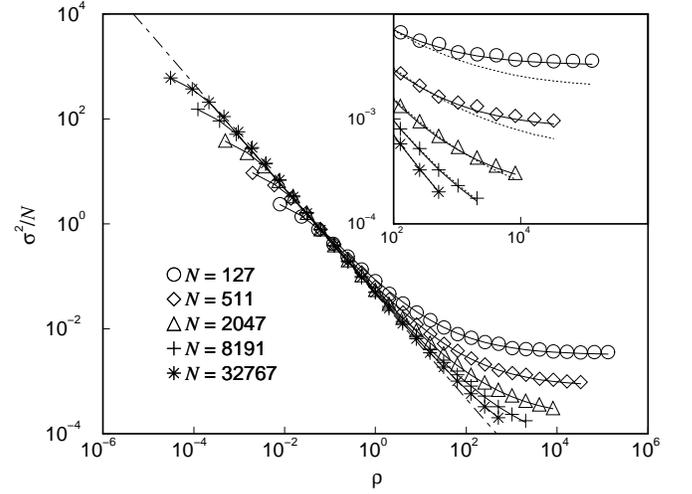,width=1\linewidth}}
\vspace{-0.4cm}
\caption{\label{vr} The dependence of the variance of the population 
for decision + 
on the diversity at $m=1$ and $s=2$.
Symbols: simulation results averaged over 1024 samples. 
Lines: theory.
Dashed-dotted line: scaling prediction. 
Inset: Comparison between simulation results (symbols),
theory with waiting effects included (lines) 
and excluded (dashed lines).}
\end{figure}

(a) {\it Multinomial regime.} When $\rho\sim N^{-1}$,
$\sigma^2/N\sim N$ with proportionality constants dependent on $m$.
To analyse this and other regimes,
we let $S_{\alpha\beta}(\omega)$ be the number of agents 
holding strategies $\alpha$ and $\beta$ (with $\alpha < \beta$),
and the virtual point of strategy $\alpha$ is initially displaced 
by $\omega$ with respect to $\beta$.
The average of $S_{\alpha\beta}(\omega)$ over initial condition is
proportional to the binomial distribution of virtual points, i.e.,
$\langle S_{\alpha\beta}(\omega)\rangle=NC^R_{(R-\omega)/2}/2^{2D-1+R}$.
The key to analysing the system dynamics 
is the observation that the virtual points of a strategy 
displace by exactly the same amount for all agents.
Hence for a given strategy pair,
the profile of the virtual point distribution remains unchanged,
but the peak position shifts with the game dynamics.
If the virtual point displacement of strategy $\alpha$ 
at time $t$ is $\Omega_\alpha(t)$,
then the agents holding strategies $\alpha$ and $\beta$
make decisions according to strategy $\alpha$ if 
$\omega+\Omega_\alpha(t)-\Omega_\beta(t)>0$,
and strategy $\beta$ otherwise.
At time $t$, we can write 
$\Omega_\alpha(t)=\sum_\mu k_\mu(t)\xi_\alpha^\mu$,
where $k_\mu(t)$ is the number of wins minus losses of decision $+$ 
up to time $t$ 
when the game responded to history $\mu$.
Consider the difference
$A^\mu(t)-A^\mu(0)=\frac{1}{N}\sum_{\alpha<\beta}S_{\alpha\beta}(\omega)
(\xi_\alpha^\mu-\xi_\beta^\mu)[
\theta(\omega+\Omega_\alpha(t)-\Omega_\beta(t))-\theta(\omega)]$.
Its average can be found by introducing the average 
$\langle S_{\alpha\beta}(\omega)\rangle$,
writing the step function as a sum over Kronecka deltas 
and introducing their integral representation,
using the identity 
$e^{ik\theta(\xi_\alpha^\mu-\xi_\beta^\mu)}
=\cos^2k\theta+i\sin k\theta \cos\theta(\xi_\alpha^\mu-\xi_\beta^\mu)
+\sin^2k\theta\xi_\alpha^\mu\xi_\beta^\mu$,
and noting that $\sum_\alpha\xi_\alpha^\mu=0$.
The final result is
\begin{eqnarray}
	\langle A^\mu(t)-A^\mu(0)\rangle\label{dA}
        &=&\frac{1}{2\pi}\int_{0}^{2\pi}d\theta
             \cos^R\theta\frac{\sin k_\mu(t)\theta}{\sin\theta}\nonumber\\
        & &\times\cos k_\mu(t)\theta\prod_{\nu\neq \mu}\cos^2k_\nu(t)\theta.
\end{eqnarray}
When $\rho\sim N^{-1}$, 
$\langle A^\mu(t+1)-A^\mu(t)\rangle\sim O(1)$  
and is self-averaging.
Since $A^\mu(0)$ is Gaussian with variance $N^{-1}$,
the values of $A^\mu(t)$ at the attractors can be computed,
and the variance found.
For example, for $m=1$, 
$\sigma^2/N
\equiv N\langle[A^{\mu^*(t)}(t)-\langle A^{\mu^*(t)}(t)\rangle]^2\rangle/4
=N[7(c_{R+1})^2-2c_{R+1}c_{R+3}+7(c_{R+3})^2]$,
where $c_n=2^{-n}C^n_{n/2}$ for even integer $n$.

(b) {\it Scaling regime.} When $\rho\sim 1$,
$\sigma^2/N\sim \rho^{-1}$ with proportionality constants 
effectively independent of $m$ for $m$ not too large.
In this case,
Eq. (\ref{dA}) can be simplified to 
$\langle A^\mu(t)-A^\mu(0)\rangle=k_\mu(t)\sqrt{2/\pi R}$.
The average step size becomes 
$\langle A^\mu(t+1)-A^\mu(t)\rangle=\sqrt{2/\pi R}\delta_{\mu\mu^*(t)}
\sim O(N^{-\frac{1}{2}})$
and is self-averaging.
To interpret this result,
we note that changes in $A^\mu(t)$
are only contributed by {\it fickle} agents
with marginal preferences of their strategies.
That is, those with
$\omega+\Omega_\alpha(t)-\Omega_\beta(t)=\pm 1$
and
$\xi_\alpha^\mu-\xi_\beta^\mu=\mp2{\rm sgn}A^\mu(t)$ for $\mu=\mu^*(t)$.
For large $R$,
the binomial virtual point distribution
among agents of a given strategy pair
is effectively a Gaussian with variance $R$.
Hence the number of agents switching strategies at time $t$ 
scales as the height of the Gaussian distribution,
which is $\sqrt{2/\pi R}$.
Thus, by spreading the virtual point distribution,
diversity reduces the step size and hence maladaptation.

As a result, 
each state of the attractor is confined in a $D$-dimensional hypercube 
of size $\sqrt{2/\pi R}$,
irrespective of the initial position of the $A^\mu$ components.
Starting from the initial state $A^\mu(0)$,
the state changes in steps of size $\sqrt{2/\pi R}$
until it reaches the attractor,
whose $2D$ historical states are given by 
$\sqrt{2/\pi R}\lceil\sqrt{\pi R/2}A^\mu(0)\rceil$ and 
$\sqrt{2/\pi R}\{\lceil\sqrt{\pi R/2}A^\mu(0)\rceil-1\}$,
where $\lceil x\rceil$ represents the decimal part of $x$.
Averaging over $A^\mu(0)$,
which are Gaussian numbers with mean 0 and variance $1/N$,
the variance of decisions become 
$\sigma^2/N=f(\rho)/2\pi\rho$,
where $f(\rho)$ approaches $(1-1/4D)/3$ for $\rho\gg 1$.
Note that $f(\rho)$ is a smooth function of $\rho$,
since $\sigma^2/N$ depends on $\rho$ 
mainly through the step size factor $1/2\pi\rho$,
whereas $f(\rho)$ merely provides a higher order correction 
to the functional dependence.
This accounts for the scaling regime in Fig.~\ref{vr}. 
Furthermore, we note that $f(\rho)$ rapidly approaches $1/3$ 
when $m$ increases.
Hence for general values of $m$,
$\sigma^2/N\rightarrow 1/6\pi\rho$, 
provided that $m$ is not too large.

(c) {\it Kinetic sampling regime.} When $\rho\sim N$,
$\sigma^2/N$ deviates above the scaling with $\rho^{-1}$,
and is given by $\sigma^2/N=f_m(\Delta)/N$,
where $\Delta\equiv\sqrt{2N/\pi\rho}$ is the {\it kinetic step size},
and $f_m$ is a function dependent on the memory size $m$.
Here $A^\mu(t+1)-A^\mu(t)$ scales as $N^{-1}$
and is no longer self-averaging.
Rather, it is equal to $2/N$ times 
the number of agents who switch strategies at time $t$,
which is Poisson distributed with a mean of $\sqrt{2/\pi R}$.
However, since the attractor is formed by steps 
which {\it reverse the sign of} $A^\mu$,
the average step size in the attractor is {\it larger} 
than that in the transient state.
To see this,
we consider the probability of $P_{\rm att}(\Delta{\bf A})$ 
of step sizes $\Delta{\bf A}$ in the attractor.
Assuming that all states of the phase space are equally likely to be accessed, 
we have $P_{\rm att}(\Delta{\bf A})=\sum_{\bf A}P_{\rm att}(\Delta{\bf A},{\bf A})$, 
where $P_{\rm att}(\Delta{\bf A},{\bf A})$ 
is the probability of finding the position ${\bf A}$ 
with displacement $\Delta{\bf A}$ in the attractor.
Consider the example of $m=1$ in the inset of Fig. \ref{va}.
The sign reversal condition implies that 
$P_{\rm att}(\Delta{\bf A},{\bf A})=P_{\rm Poi}(\Delta{\bf A})
\prod_\mu\theta[-A^\mu(A^\mu+\Delta A^\mu)]$,
where $P_{\rm Poi}(\Delta{\bf A})$ is the Poisson distribution of step sizes,
yielding $P_{\rm att}(\Delta{\bf A})=P_{\rm Poi}(\Delta{\bf A})\prod_\mu\Delta A^\mu$.
Thus the attractor averages $\langle(\Delta A^\pm)^2\rangle_{\rm att}$,
which are required for computing the variance of decisions, are given by
$\langle(\Delta A^\pm)^2\Delta A^+\Delta A^-\rangle_{\rm Poi}/
\langle\Delta A^+\Delta A^-\rangle_{\rm Poi}$.
In other words, the sampling of the step sizes 
is weighted by the attractor sizes due to the kinetics.
The result for $m=1$ is 
$\sigma^2/N=(14\Delta^3+105\Delta^2+132\Delta+24)/96N(2\Delta+1)$.

(d) {\it Waiting regime.} When $\rho\gg N$, 
$\sigma^2/N$ deviates above the predictions of kinetic sampling.
Here the agents are so diverse 
that the average step size is approaching 0. 
At each state in the phase space, 
the system remains stationary for many time steps, 
waiting for some agent to reduce the magnitude of her virtual point 
until strategy switching can take place. 
This waiting effect modifies the composition of the group of fickle agents 
who contribute to the state transitions, 
and consequently increase the step sizes and variance 
above those predicted by kinetic sampling. 
Consider the example of $m=1$. 
As shown in the inset of Fig. \ref{va}, 
the attractor consists of both vertical and horizontal hops, 
and detailed analysis shows that 
only one type of agents can complete both hops.
Since fewer and fewer agents contribute to the switching of states 
in the limit $\rho\gg N$, 
a single agent of this type will dominate the game dynamics, 
and one would expect that $\sigma^2/N$ approaches $1/4N$. 
However, when waiting is possible, 
agents not of this correct type can wait for other agents 
to complete the hops in the attractor, 
even though one would expect that 
the probability of finding more than one fickle agents 
is drastically less than that for one. 
In fact, analysis shows that the attractor consists of a single fickle agent 
with a probability of 1/11 only, 
and $\sigma^2/N$ approaches $9/22N$ rather than $1/4N$. 
As shown in the inset of Fig.~\ref{vr},
lengthy analytic results including waiting effects significantly improve 
the agreement with simulations over the kinetic sampling prediction.

Many properties of the system dependent on the transient dynamics 
also depend on its diversity.
For example, since diversity reduces the fraction of agents 
switching strategies at each time step,
it also slows down the convergence to the steady state.
Hence in the scaling regime, 
the convergence time scales as $\rho^{1/2}$.
Specifically, when $\rho\gg 1$,
the average convergence time becomes $(2+\sqrt{2})\sqrt{\rho}$ for $m=1$.
Similarly, the distribution of payoffs
among the {\it frozen} agents
(that is, agents who do not switch their strategies at the steady state)
also depends on the transient.
Since the system dynamics reaches a periodic attractor,
they have constant average payoffs at the steady state.
Hence any spread in their payoff distribution 
is a consequence of the transient dynamics.
Thus, in the scaling regime, 
the mean square payoff scales as $\rho$.
Specifically, when $\rho\gg 1$,
the mean square payoff becomes $\pi\rho$ for $m=1$.
Simulation results of both the convergence time and the mean square payoff
have an excellent agreement with the theory \cite{elsewhere}.

The results presented here can be generalized to other cases.
Consider the {\it exogenous} MG,
in which the information $\mu(t)$ was randomly 
and independently drawn at each time step $t$ \cite{statmech}.
The picture that the states of the game are hopping between hypercubes 
in the phase space remains valid.
At the steady state,
the attractor consists of hoppings among all vertices of a hyperpolygon 
enclosing the origin in the phase space,
analogous to the present {\it endogenous} case,
in which a fraction of hyperpolygon vertices belong to the attractor.
In the scaling regime,
the behavior depends on the scaling of the step sizes with diversity,
rather than the actual sequence of the steps.
Consequently,
the behavior is similar to that of the endogenous game.

The present results can be extended to higher values of $m$ \cite{elsewhere}. 
For $m=2$, analysis using the de Bruijn sequence explicitly 
yields excellent results. 
For higher $m$, we approximate the attractor of the exogenous game 
by a hyperpolygon enclosing the origin of the phase space.
Using a generating function approach,
and taking into account the scaling of step sizes and kinetic sampling, 
the computed variance of decisions 
agrees qualitatively with simulations, 
except for values of $\alpha$ close to $\alpha_c$.

We can also make qualitative predictions
about the transition from the symmetric to asymmetric phase
when the complexity $\alpha$ increases \cite{symmetry}.
From Eq. (\ref{dA}),
the average displacement in the phase space is given by
\begin{eqnarray}
        \langle A^\mu(t)-A^\mu(0)\rangle\approx
	k_\mu(t)\sqrt{\frac{2}{\pi(R+2D\langle k^2\rangle)}}, 
\label{dA2}
\end{eqnarray}
where $\langle k^2\rangle$ represents the mean of $k_\nu(t)^2$ 
for all $\nu\leq D$.
For $\rho\sim\alpha\sim 1$, 
it can be verified that $A^\mu(t)-A^\mu(0)$ is self-averaging. 
Suppose the game dynamics leads to an attractor near the origin,
with $\langle A^\mu(t)\rangle\to 0$.
Noting that $\langle A^\mu(0)^2\rangle\sim 1/N$,
we obtain the self-consistent relation 
$\langle k^2\rangle=\rho/2(\alpha_c-\alpha)$,
where $\alpha_c=1/\pi\approx 0.318$.
This means that when $\alpha$ approaches $\alpha_c$,
the average step size appraoches 0 in the asymptotic limit.
There is a critical slow down since the convergence time diverges.
When $\alpha$ exceeds $\alpha_c$,
the average step size vanishes before the system 
reaches the attractor near the origin,
so that the state of the system is trapped at locations 
with at least some components being nonzero.
The interpretation is that when $\alpha$ is large, 
the distributions of strategies become so sparse that 
motions in the phase space cannot be achieved by the switching of strategies.
Note that the value of $\alpha_c$ is close to the value of 0.337 
obtained by the continuum approximation \cite{statmech} 
or batch update \cite{coolen}.

From the perspective of game theory,
it is natural to consider whether the introduction of diversity 
assists the game to reach a Nash equilibrium.
It has been verified that Nash equilibria consist of pure strategies 
\cite{statmech}.
Hence all frozen agents have no incentives to switch their strategies.
In fact, since the dynamics in the attractor is periodic,
the payoffs of all strategies become zero when averaged over a period.
Thus, the Nash equilibrium is approached in the sense that 
the fraction of fickle agents decreases with increasing diversity.
In the extremely diverse limit,
it is probable that only one fickle agent
switches strategy at each step in the attractor.
In this case, even the fickle agent cannot increase her payoff,
since on switching she always remains on the majority side and loses.
Then a Nash equilibrium is reached exactly.
For $m=1$, for example, a Nash equilibrium is reached in this way 
with probability $7/11$.

In summary, we have studied the effects of diversity 
in the initial preference of strategies 
on a game with adaptive agents competing for finite resources.
Scaling of step sizes accounts
for the behavior of the variance of decisions
in the scaling regime ($\rho\sim 1$).
At high diversity, 
we find that the scaling mechanism is supplemented by kinetic sampling,
a mechanism self-imposed by the requirement to stay in the attractor.
In extremely diverse systems, 
we discover further a waiting mechanism, 
when agents who are unable to complete the attractor dynamics alone 
wait for other agents to collaborate with them.
Together, these mechanisms yield theoretical predictions
with excellent agreement
with simulations over 9 decades of data.
By introducing diversity,
the variance of decisions in the symmetric phase decreases,
showing that the maladaptive behavior is reduced.

The combination of scaling, kinetic sampling and waiting
in accounting for the steady state properties of the system 
illustrates the importance of dynamical considerations 
in describing the system behavior. 
We anticipate that these dynamical effects will play a crucial role 
in explaining the system behavior in the entire symmetric phase, 
since when $\alpha$ increases, 
the state motion in a high dimensional phase space 
can easily shift the tail of the virtual point distributions 
to the verge of strategy switching, 
leading to the sparseness condition 
where kinetic sampling and waiting effects are relevant. 
Due to the generic nature of these effects,
we expect that they are relevant 
to minority games with different payoff functions and updating rules, 
as well as other multi-agent systems. 

We thank P. Luo, Y. C. Zhang, L. H. Tang, B. H. Wang
for fruitful discussions.
This work is supported by the research grants HKUST6153/01P and HKUST6062/02P
from the Research Grant Council of Hong Kong.

\bibliography{scaling7}

\end{document}